\documentclass[a4paper,twosides]{article}
\usepackage{multicol,multirow,graphics,fancyhdr,epstopdf}
\usepackage{graphicx} 
\usepackage{amsmath,amsfonts,amssymb,bm,upgreek,mathrsfs,ccmap,mathcomp}
\usepackage[pagewise,switch,columnwise]{lineno}
\usepackage[compress,nospace]{cite}
\usepackage[dvipsnames]{xcolor}
\usepackage{CPL-2022}
\newcommand{\cplyear}{2024} \newcommand{\cplvol}{41}
\newcommand{\cplno}{x} \newcommand{\cplpagenumber}{119701}
 \newcommand{\cplpage}{\cplpagenumber-\thepage}

\begin{document}

 \begin{center}
\large\bf{\boldmath{Observations of fast radio variations in microquasars by FAST}}
\footnotetext{\hspace*{-5.4mm}$^{*}$Corresponding authors. Email: wangwei2017@whu.edu.cn

\noindent\copyright\,{\cplyear}
\href{http://www.cps-net.org.cn}{Chinese Physical Society} and
\href{http://www.iop.org}{IOP Publishing Ltd}}
\\[5mm]
\normalsize \rm{}Wei Wang$^{*}$
\\[3mm]\small\sl Department of Astronomy, School of Physics and Technology, Wuhan University, Wuhan 430072, China


\normalsize \rm{}(Received xxx; accepted manuscript online xxx)
\end{center}

\vskip 1.5mm

\small{\narrower Microquasars are the compact objects generally including accreting black holes which produce relativistic jets. The physical mechanisms of jet launching, collimation, and acceleration are poorly understood. Microquasars show strong variability in multi-wavelength observations. In X-rays, the sources show the fast variation features down to millisecond time scales, with the prominent quasiperiodic oscillations (QPOs) around 0.1 Hz - tens of Hz in light curves, however, physical origin of QPOs is still uncertain. FAST as the largest radio telescope provides the opportunity to study fast variability of both radio flux and polarization in microquasars. In the FAST observations from 2020 - 2022, we reported the first evidence of radio subsecond quasi-periodic oscillations of GRS 1915+105, providing the direct link between QPOs and the dynamics of relativistic jets. These QPOs with the centroid frequency around 5 Hz are transient, accompanied with strong evolution of the spectral index. Combined with multiwavelength observations, we discuss the possible physical models to produce radio QPOs in BH systems: the helical motion of jet knots or precession of the jet base. In near future, high time resolution radio monitoring of microquasars based on FAST is expected to discover more new phenomena in black hole systems, which will be important to understand the physics in strong gravity. 

\par}\vskip 3mm
\normalsize\noindent{\narrower{PACS: xxxx}}\\
\noindent{\narrower{DOI: \href{http://dx.doi.org/10.1088/0256-307X/\cplvol/\cplno/\cplpagenumber}{10.1088/0256-307X/\cplvol/\cplno/\cplpagenumber}}

\par}\vskip 5mm
Astrophysical black holes (BHs) mainly include the stellar mass BHs in the X-ray binaries, and supermassive BHs in the center of galaxies. The supermassive BHs accrete the materials from the central regions of the Galaxies, and produce very luminous compact emitting sources in distant universe, with relativistic jets observed in these systems, which are generally named as active galactic nuclei or quasars.\ucite{1} X-ray binaries generally including a stellar mass BH and a normal star, are one of the brightest X-ray sources in the Galaxy. In these systems, the BH accrete the material from the companion star, produce X-rays in the accretion disk, and radio emissions from relativistic jets, therefore, also named as microquasars.\ucite{2,3}

Microquasars and quasars have similar scenarios for accretion processes, relativistic jets, radiation mechanisms except for the quite different mass ranges for the central black holes (see Figure 1).\ucite{4,5,6} The radiations from these BH systems generally show the variations, and the typical timescales of emission variations would be connect to dynamical scales which can be characterized by the inner radius of accretion disk. The inner radius should be several times of the Schwarzschild radius of the BH which is depending on the BH mass: $R_{S}=2GM/c^2$. Thus, the quasars and microquasars have different BH masses, then the typical emission sizes are different, which leads to the different variation timescales in two BH systems. For quasars, the emission size will be around $l\sim 10^{9-10}$ km,\ucite{4,5} the typical variation timescale given by $\delta\sim l/c$ is about hours to days. However, the inner accretion disk radius in microquasars is about $10^{2-3}$ km,\ucite{6} which results in the minimum variation timescales to $\sim 10$ ms. Then timing analysis of microquasars will require the high time-resolution observations to probe the plasma dynamics near the BH.

Microquasars behave as the black hole X-ray binaries,\ucite{3,6} which have been frequently observed by the space X-ray telescopes which can record the arrival time of each X-ray photon with a time resolution below 1 ms. In fast X-ray variation patterns, there exists a special feature observed in all BH X-ray binaries, i.e. quasi-periodic oscillations (QPOs), manifesting in the power spectrum as narrow, harmonically related peaks.\ucite{7} With different X-ray missions, QPOs have different frequencies: low-frequency QPOs (LFQPOs) with the typical frequency range of $\sim 0.1-10$ Hz are observed in nearly all BH systems,\ucite{8,9,10,11} and high-frequency QPOs (HFQPOs) with the frequency range of $\sim 30-100$ Hz are only observed in limited sources.\ucite{12,13} With the Hard X-ray Modulation Telescope (\textit{Insight-}HXMT) covering a wide X-ray band from 1 -250 keV, fruitful information of QPOs is discovered, e.g., high energy QPOs,\ucite{14,15,16,17} transient QPOs,\ucite{18,19,20} which help to understand the physics origin of QPOs and the parameters of BH systems.\ucite{21,22} However, the production mechanism of QPOs is unknown,\ucite{7} and only X-ray observations cannot justify which component would contribute to flux oscillations: the inner accretion disk, corona or outflow. There exist different theoretical models to explain the special features, e.g., accretion ejection instability,\ucite{23} propagating oscillatory shocks in the disk,\ucite{24} relativistic precession of the inner accretion flow or jet base.\ucite{25,26}

\vskip 4mm

\fl{1}\includegraphics[width=0.9\columnwidth]{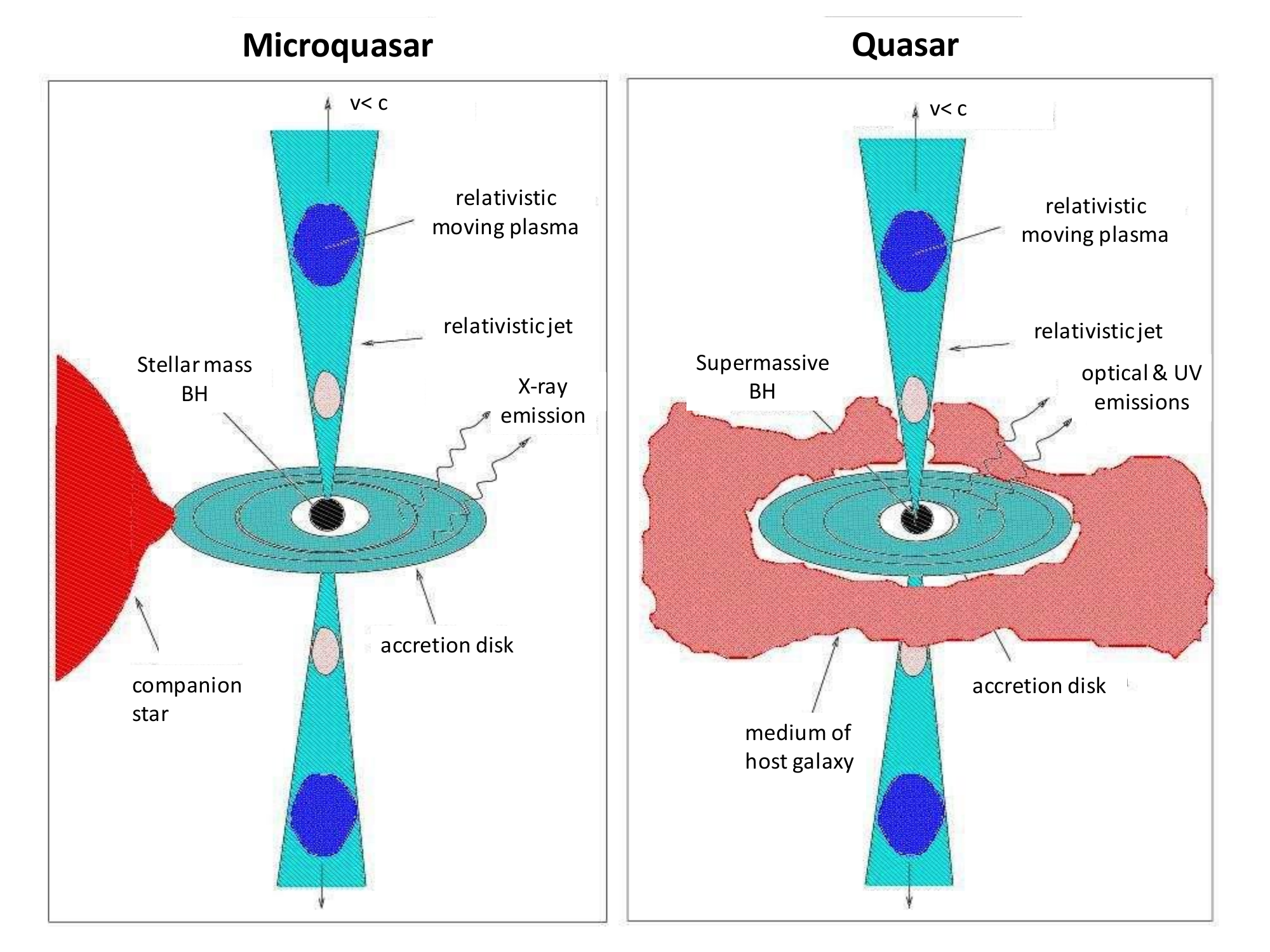}
\vskip 2mm
\figcaption{7.5}{1}{The cartoon figures show the comparison of two black hole accreting systems: microquasars versus quasars. Quasars host the supermassive BHs with the masses ranging from $\sim 10^6-10^{10} \ M_\odot$ in the center of the galaxies,\ucite{5} which accrete the materials from the medium in the galaxy, and produce optical, ultraviolet (UV) and X-ray emissions. Microquasars are located in a binary system with a stellar mass BH (e.g., $5 - 30\ M_\odot$) accreting from the material from the companion star,\ucite{3} and forming the accretion disk which dominantly emits X-rays. Both BH systems can launch relativistic jets which would produce the non-thermal emissions in radio or gamma-ray bands. }

\medskip

Radio observations of microquasars can directly study the launching, dynamics and radiation mechanism of relativistic jets.\ucite{1,2} At present, there are two traditional observation patterns of microquasars in radio bands. One is the imaging observation by interference, like VLBI and VLA, and the interference observations have the advantage of very high angular resolution limit ($\sim 10^{-4}$ arcsec),\ucite{27} which provide the clear pictures for the moving knots of relativistic jets.\ucite{28,29} However, the present interference observations have the timing limit up to 5- 10 s.\ucite{30,31} The other way to study radio variations of the jets is the long-term monitoring of the BHs using single aperture telescopes, due to the limit of sensitivity of the previous radio telescopes, the observations give each data point per several minutes/hours.\ucite{32,33} Thus during the multiwavelength studies of microquasars, X-rays can study the variations down to $\sim 1$ millisecond,\ucite{34} but radio bands have the timing resolution larger than 5 seconds, then the differences make it impossible that the variation studies with different bands can be performed in the similar timescale (e.g. $<1$ s).\ucite{31} Then if we request the high-resolution timing studies of microquasars, the more sensitivity radio observations are needed.

The FAST telescope's total and effective apertures are 500 and 300 meters, respectively, then the sensitivity is more than ten times better than other present radio telescopes.\ucite{35} So FAST provides us the good chance to perform the high-resolution timing radio observations of microquasars, which may probe the fine structure in variations of relativistic jets. Fast variation patterns in microquasars are important to understand the accretion process and plasma dynamics near the stellar-mass black holes. In the design of FAST, there are two main science projects: pulsars and hydrogen spectral lines. Microquasars would be the new interesting and unique field based on high sensitivity observations of FAST.

In this Letter, we will at first present the detailed description of our idea to perform the high-resolution timing observations of microquasars, data processes and the discovery of radio subsecond QPOs with FAST observations from 2020 - 2022. The implications of the early discovery will be also elucidated. Finally the future observation plans and prospective for the BH physics are briefly discussed.

{\it Methods and data processes.} 
FAST has several observation modes with the 19-beam feed receiver covering the band of 1.0-1.5 GHz, and we performed several continuous high-resolution timing observations of a famous microquasar GRS 1915+105 with tracking and on-off modes based on our scientific purposes from 2020 - 2022. The observations have used the central beam (M01) and other 18 beams with tracking and on-off modes, and before and after the target observations, the on-off source measurements are performed for the calibration. The field of view (FOV) of M01's single beam is around 2.9$^\prime$ at 1.4 GHz. The data stream has 128 subints, 1024 spectra, complete Stokes information, and 4096 channels. The channel width and time resolution are 0.1220703125 MHz and 98.304 microseconds, respectively.\ucite{35}

The data of FAST are recorded in PSRFITS format.\ucite{36} For each of FITS file, we do the re-sampling for the original data and extract the frequency band and time from 4096 frequency channels and 128 subints, and then combine the re-sampled preprocessed data files. The PRESTO would produce a time series which is the uncalibrated lightcurve from the combined file.\ucite{36} In the followings, we briefly introduce the data processes of the radio frequency interference (RFI) elimination, calibration of the total intensity and polarization.

The elimination of radio frequency interference (RFI) is very important for the radio data processes. For our work, we use {\tt PRESTO's rfifind} software \ucite{36} based on the fast Fourier transform algorithm to find the RFI over a window of 0.1 seconds interval and create a mask to eliminate these narrow-band noises. Figure 2 shows an example of a mask using {\tt rfifind} that marks not just broadband noises but also blob RFI. For multi-frequency data reduction, we manually selected the mask, then split the remaining channels into numerous segments and averaged them individually, that procedure would therefore reduce the affection of RFI, so that we could get rather clean channels for further data reduction.\ucite{37}

After identifying the RFI and producing the masks, we would use the pulsar search and analysis software {\tt PRESTO} to generate the dedispersed time series which can be used for further calibration. The observed radio flux density with time can be calculated by
\begin{equation}
\label{eq:caltot}
\begin{split}
    Flux(t)= \frac{I(t)-OFF}{ON-OFF}\cdot T_{src}\cdot\frac{1}{G},
\end{split}
\end{equation}
where $I(t)$ is the observed intensity value from the telescope with time in the tracking mode, $T_{src}$ is the temperature of the telescope due to the astronomical source, $ON$ and $OFF$ are the intensity values in the turning-off state of the noise diode when the telescope is directed at the celestial source and background sky in on-off mode respectively, and $G$ is the full gain of FAST in the sky coverage. Then we can determine the temperature of the telescope when it is gazing at the source and the background sky by using data from the on-off mode recorded before the begin and after the end of observation session. This calibration method would yield the relative, rather than the absolute flux density for one observation. If we hope to get the absolute flux level, we will observe the calibration sources (e.g., 3C 286) before the target observations.

\vskip 4mm
\begin{center}
\fl{1}\includegraphics[width=0.76\columnwidth]{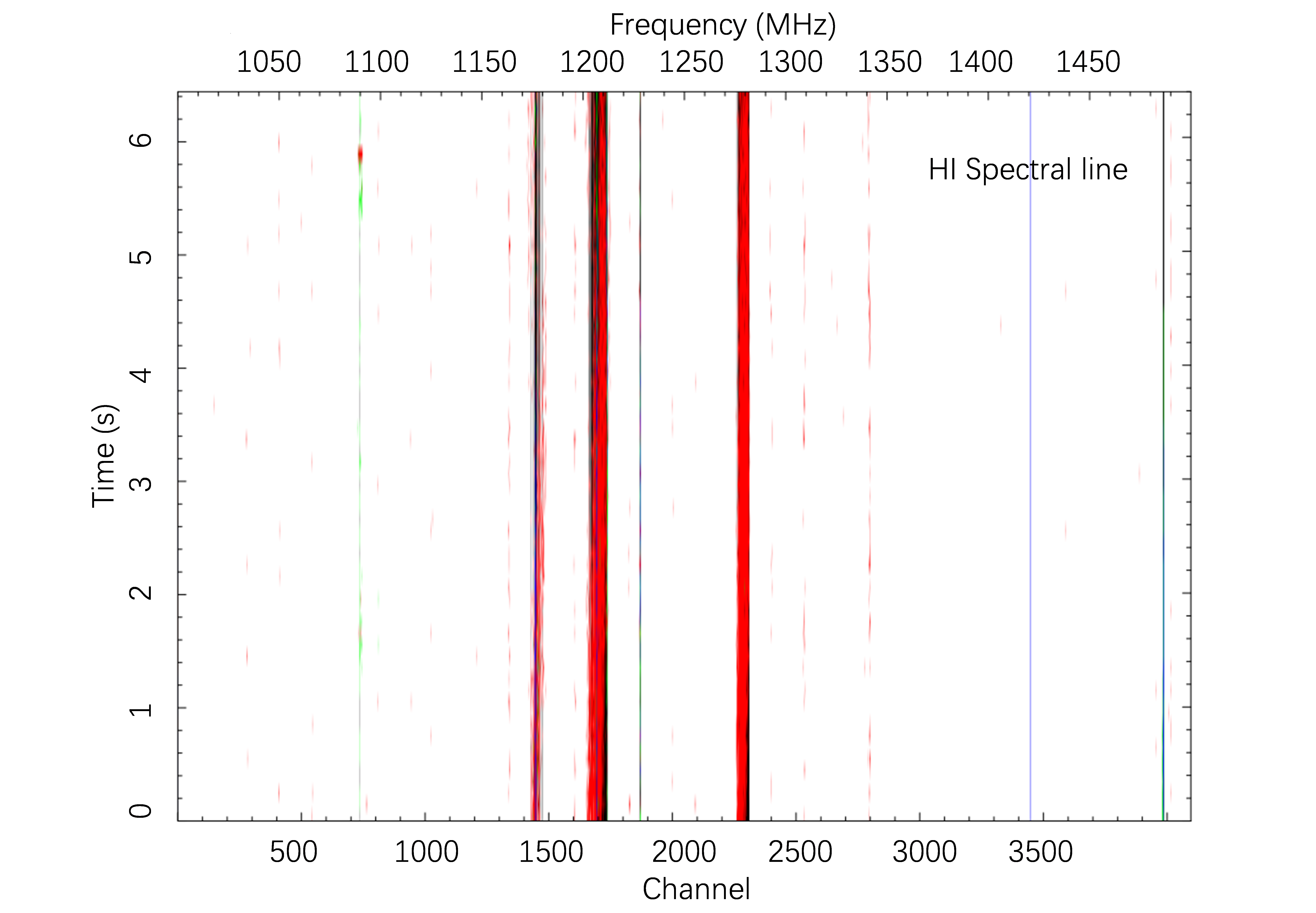}
\end{center}
\vskip 2mm
\figcaption{7.5}{2}{A time-frequency image of Mask generated via one {\tt PSRFITS} file by software {\tt rfifind} presents the RFIs.\ucite{37} The X-axis represents frequency (channel) range from 1000-1500 MHz and the Y-axis represents time for 6.4 seconds. The most common powerful RFIs ranging from 1150 to 1300 MHz come from satellites and flagged by vertical colored strips. }

\medskip

The full polarization data are recorded by FAST with standard FITS-based format for pulsar data files ({\tt PSRFITS}),\ucite{36} which contain the polarization components which are recorded as the AABBCRCI form, where AA and BB are the direct products of two channels, CR and CI are the real and imaginary parts of the cross product of two channels respectively. We fold the original files with their duration as folding periods to produce the four time series with {\tt DSPSR}.\ucite{38} Then we should eliminate the RFIs for each time series. After RFI removal, we used {\tt PSRCHIVE} \ucite{39} to extract four coherence data: $I_x^2$, $I_y^2$, CR and CI, where CR and CI are the real and imaginary parts of the cross product $I_x*I_y$. Feeds in radio telescope are not perfect, so we should consider the relative gain of electronics system (defined as leakage $f$) and phase differences between two channels (phase error $\delta_{er}$) for precise polarization calibration, and M\"uller matrix of the radio feeds was taken into account to amend the Stokes parameters from observations.\ucite{40}

The leakage $f$ is defined as $f=\frac{Q_{obs}^{\prime}}{I_{obs}^{\prime}}$, and the phase error as $\delta_{er}=\frac{1}{2}\arctan\frac{V_{obs}^{\prime}}{U_{obs}^{\prime}}$, where Stokes parameters with subscript $obs$ and $true$ refer to before and after M\"eller matrix calibration, the prime means noise diode signal. Them we can get the true values of the four Stokes parameters:\ucite{37}

\begin{equation}
\begin{aligned}
    I_{true}&=\frac{I_{obs}-f*Q_{obs}}{1-f^2}\\
    Q_{true}&=\frac{Q_{obs}-f*I_{obs}}{1-f^2}\\
    U_{true}&=P\cos[2(\delta-\delta_{er})]\\
            &=U_{obs}\cos2\delta_{er}
             +V_{obs}\sin2\delta_{er}\\
    V_{true}&=Psin[2(\delta-\delta_{er})]\\
            &=V_{obs}\cos2\delta_{er}
             -U_{obs}\sin2\delta_{er},
\end{aligned}
\end{equation}
where $P=\sqrt{U_{obs}^2+V_{obs}^2}$ is the root square of $U_{obs}$ and $V_{obs}$, and $\delta$ is the phase between $U_{obs}$ and $V_{obs}$. Finally we can calculate the linear, circular polarization and polarization position angle from calibrated Stokes parameters,
\begin{equation}
\begin{aligned}
    LP&=\frac{\sqrt{Q_{true}^2+U_{true}^2}}{I_{true}}\\
    CP&=\frac{V_{true}}{I_{true}}\\
    PA&=\frac{1}{2}\arctan\frac{U_{true}}{Q_{true}}.
\end{aligned}
\end{equation}

{\it Early science results:} 
The first science objective GRS 1915+105 is a massive BH system in the Galaxy,\ucite{41} containing a fast-spinning black hole ($a^* >0.99$).\ucite{42} The superluminal motion of radio emission from this source indicates relativistic jets of velocity higher than 0.9 c.\ucite{43} It exhibits persistent X-ray activity over the last 30 years, with quasi-periodic oscillations (QPOs) of $\sim 1-9$ Hz,\ucite{44,45} and $\sim$ 34 and 67 Hz in the X-ray bands.\ucite{46} From 2020-2022, we perform 10 observations of GRS 1915+105 with FAST. In the early 8 observations in 2020 during the test operations of FAST, three of observations have no calibrations, and 2 of the observations have the strong artificial interferences and could not be calibrated, thus only three of observations can be used for the further studies with less noises.\ucite{37}
\vskip 4mm
\begin{center}
\fl{1}\includegraphics[width=0.3\columnwidth]{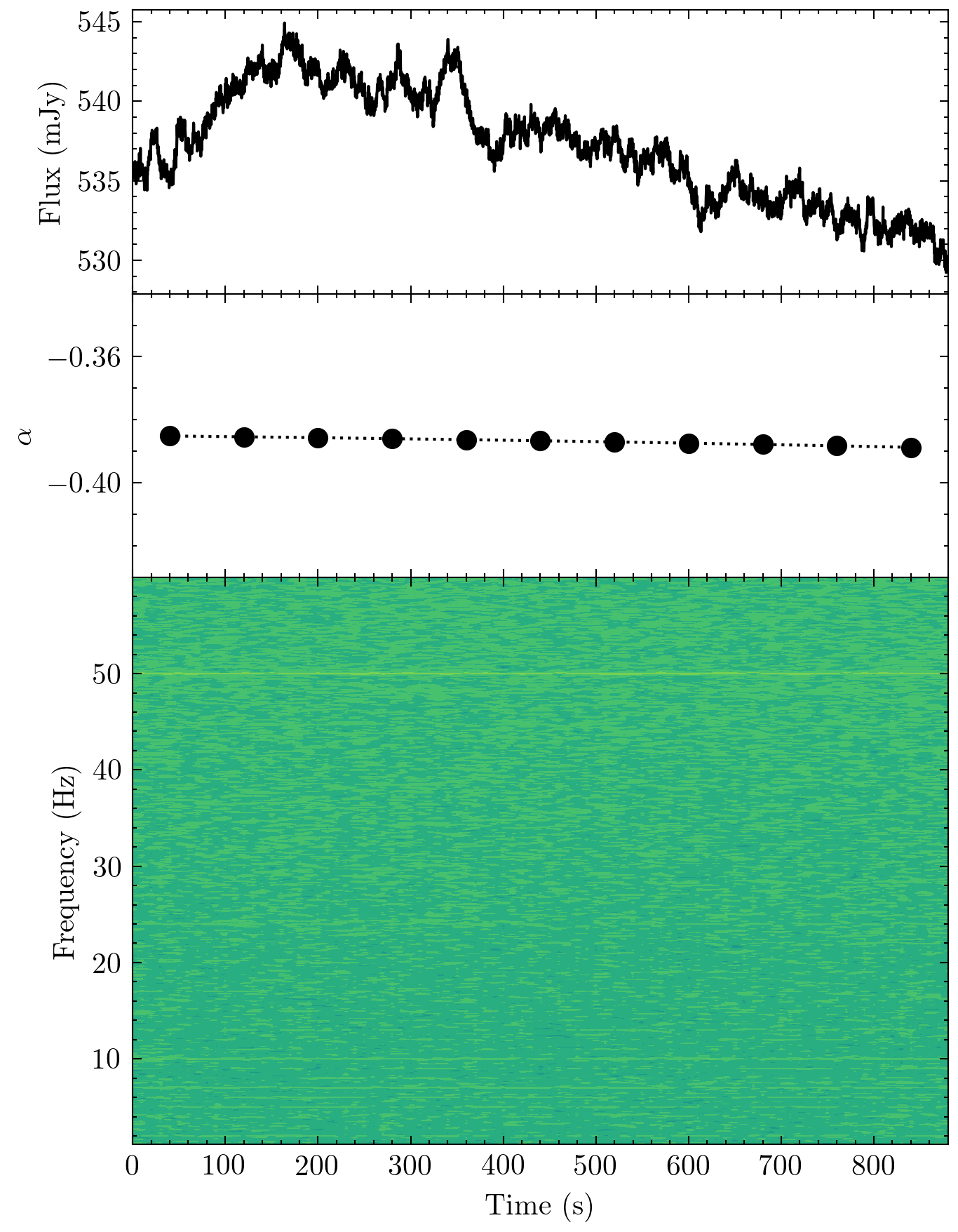}
\fl{1}\includegraphics[width=0.3\columnwidth]{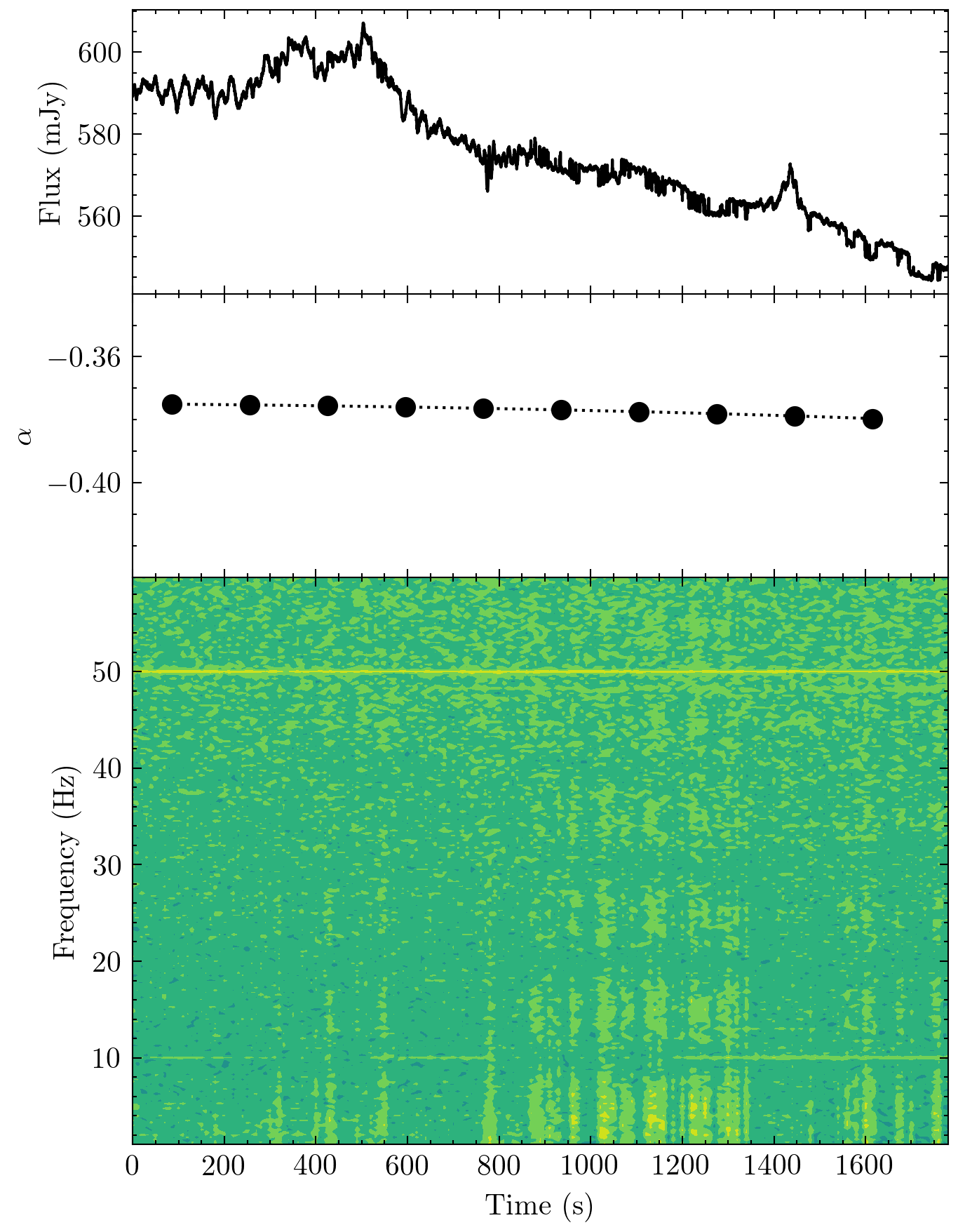}
\fl{1}\includegraphics[width=0.3\columnwidth]{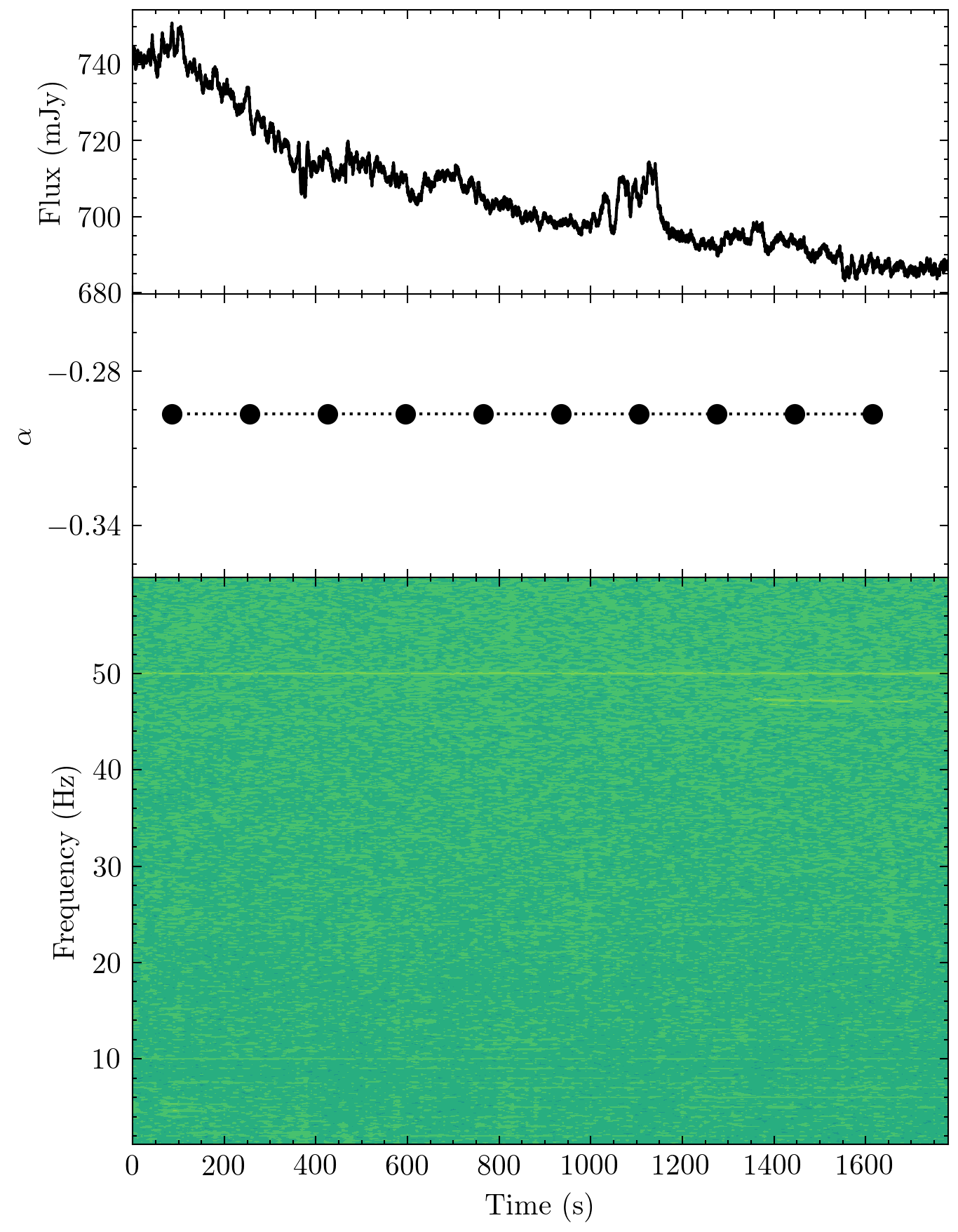}
\end{center}
\vskip 2mm
\figcaption{7.5}{3}{The radio flux density, spectral index evolution and dynamical power density spectra of the microquasar GRS 1915+105 for three FAST test observations in 2020: Feb 10 (left panel), Feb 12 (middle panel) and March 8 (right panel) 2020, respectively. During these observations, radio flux did not show any strong flares, the spectral index $\alpha$ was nearly a constant during the observations, and the dynamical PDSs had no QPO features from 0.1 - 100 Hz. The 50-Hz features in the whole time are due to the alternating current frequency. }
\medskip

To study the fast radio variations of the source, we derive the light curves for the flux density, degrees of linear polarization (LP), circular polarization (CP), and linear polarization position angle (PA) over the observing time intervals with a time resolution of $\sim$ 0.002 s after data reduction and calibration. Then we calculate the dynamical power spectra of these light curves, which aims to search for the possible quasi-periodic oscillations in radio bands. During the three observations in 2020, the flux densities of the source during each observation are relatively stable without any significant QPO signals from $\sim 0.1 - 100$ Hz (see Figure 3), the spectral index is near a constant for each observation of $\alpha\sim -(0.4 - 0.3)$ with the linear polarization between 20-30$\%$ and the polarization position angle around 89.5$^\circ$.\ucite{31} While in the observations in January of 2021 and June of 2022, with the better performance of FAST observations, we discovered subsecond QPOs. In the dynamical power spectra of radio flux density, there exist transient QPO signals around 0.2 s in GRS 1915+105 (also see the power spectra in Figure 4).\ucite{47}

During the occurring of the QPOs in January of 2021 which lasted about 1200 seconds, the total intensity flux density showed the increase trend with the significant spectral index evolving from $\sim -0.6$ to $-0.3$. This rapid changes of the radio spectral index imply the occurrence of an ejection event.\ucite{48} Before and after the QPO epoch, the flux density is stable or reaches a plateau. The linear polarization increases from $\sim 25\%$ {\bf to $31\%$,} the position angle is relatively stable around $\sim 88^\circ$. The circular polarization is near zero before and after the QPOs, while CP becomes left-handed during the QPO epoch (e.g., average CP of $-1\%$). In addition, the very weak harmonic signal around 10 Hz are detected in the 2021 observations.\ucite{49}

\vskip 4mm
\begin{center}
\fl{1}\includegraphics[width=0.65\columnwidth]{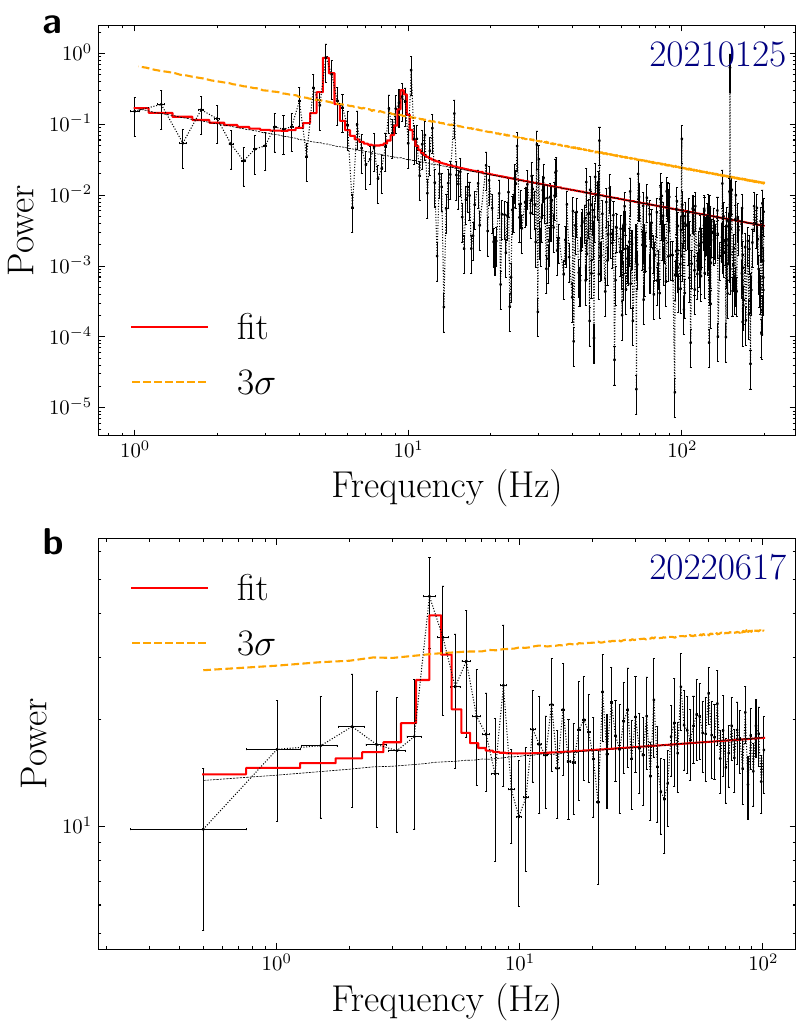}
\end{center}
\vskip 2mm
\figcaption{7.5}{4}{The average power spectra of radio flux light curves based on the FAST observational data.\ucite{47} {\bf a:} The power density spectrum (PDS) of the light curve selected from the epoch observed on 2022-01-25, and the PDS is fitted with a power-law component and two Lorentzian functions. {\bf b:} The PDS of the radio light curve observed on 2022-06-16. The sub-second QPOs around 5 Hz were reported from the target source. The power spectrum is fitted with a power-law component and a Lorentzian function. The colored dashed line in the middle panel indicates the confidence level at $3\sigma$, using a light curve simulation algorithm \ucite{50}. For the significance level computation, we simulated 20000 light curves with power-law distributed noises appropriate for our data and resampled these light curves to ensure the resolution which matched that of our observation data. }

\medskip

The transient QPO signals around 5 Hz detected in June 2022 lasted about 100 seconds. During this QPO event, the radio flux density was relatively steady at a level around 350 mJy, with the measured LP $\sim 6.5\%$, CP $\sim -1.3\%$ and PA $\sim 96^\circ$. The spectral index $\alpha$ evolved from $-0.08$ to $\sim -0.01$.\ucite{47} In addition, we also performed the multi-beam data analysis for the QPO signals in 2022. During the observations, the central beam (M01, field of view (FOV) of 3 arcmin) was toward the target GRS 1915+105, while the beams M02 -- M19 were toward other sky regions for background monitoring. Only the M01 data showed the QPO peak feature at around 5 Hz in the power spectrum, and other beam light curves only showed fluctuations in the power spectra with no QPO features detected. This detection strongly suggested that the radio QPOs come from the microquasar.

The transient sub-second quasi-periodic oscillations have the complicated temporal variation structure which could be connected to jet dynamics of the black hole (BH). To probe the fine structure of the QPO feature, we carry out wavelet analysis (detailed method descriptions refer to \ucite{51,52}) of the flux light curves during the epoch of periodic oscillations (see two examples presented in Figure 5), which can show the variations of the QPO signals and flux variations in short time scales (less than 0.1 s). In the high time-resolution wavelet power diagram the 5-Hz signal still shows a discontinuous behavior, with the QPO signal disappearing sometimes, while the 10-Hz signal is more sparse or non-detected. A statistical study of the timescales of the detected transient QPO signals based on the wavelet analysis finds the duration of the 5-Hz QPO to be about $0.4 - 12$ s with a peak of $\sim 0.7$ s.\ucite{47} The typical duration of the 5-Hz QPO signal at $\tau \sim 0.7$ s defines a characteristic scale $\sim \tau c \sim 2\times 10^{10}$ cm, $c$ is the light speed, which may be related to the typical size of the QPO emission region or lower limit of the emission height.

\vskip 4mm
\fl{1}\includegraphics[width=0.88\columnwidth]{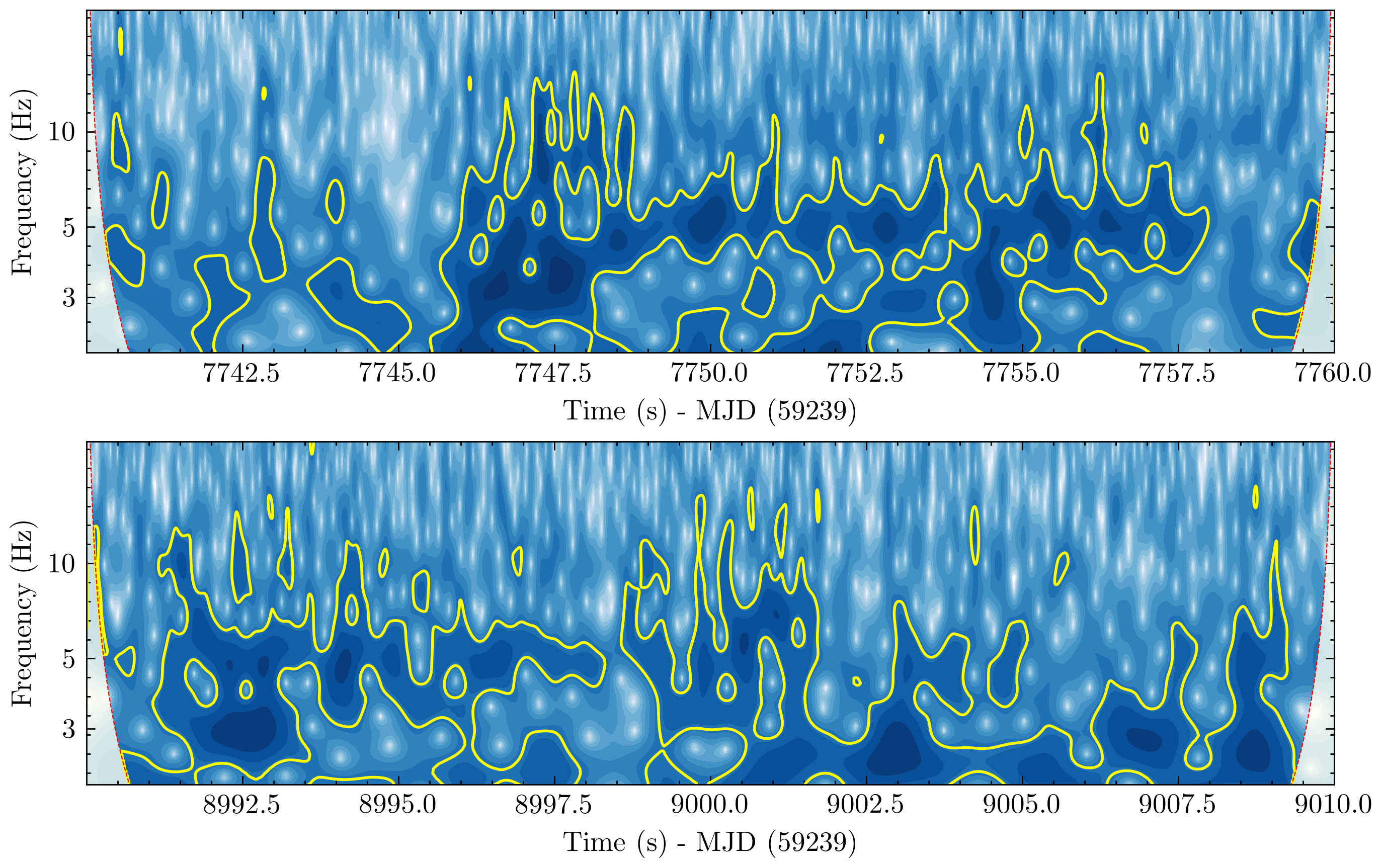}

\vskip 2mm
\figcaption{7.5}{5}{The wavelet spectra of the radio flux density for two time domains with detecting the QPOs of GRS 1915+105 in Jan 2021: the appearance of the QPO (top panel) and the disappearance of the QPO (bottom panel). The QPO signals in the wavelet spectra show the complicated structure both in the frequency and time domains, the frequency may have the turbulent shifts, and the QPO signals would appear and disappear in short time scales.  }
\medskip

{\it Physical implications:} 
The low-frequency QPOs of black hole X-ray binaries with the frequency range of $\sim 0.01-20$ Hz are observed in X-rays,\ucite{10,11,12,13,14} optical,\ucite{53} infrared\ucite{54} and ultraviolet\ucite{55} bands. Long-period radio QPOs with the periods from about one hundred days to several years have been reported in some radio loud active galactic nuclei (AGNs), specially blazars.\ucite{56,57,58,59} These radio QPOs generally last for about several to twenty cycles, which likely reflect the special dynamics of relativistic jets powered by supermassive black holes (SMBHs) in AGNs. Radio oscillations with a period of $\sim$ 15 hours was also found in a gamma-ray X-ray binary LS I$+61^\circ$303,\ucite{60} which only had two or three QPO cycles. In addition, slow radio oscillations in the period range of the 20-50 minutes were detected in GRS 1915+105.\ucite{61,62} The X-ray light curves of GRS 1915+105 revealed the low-frequency QPOs from $\sim 1-9$ Hz,\ucite{44,45} and the high-frequency QPOs at $\sim 34$ and 67 Hz.\ucite{46} The high-frequency QPOs may be the Kepler frequency of the accretion disk near the BH,\ucite{46} while the mechanism to produce the low-frequency QPOs is still in dispute.\ucite{7}

There have been a few physical models suggested to interpret these QPOs. For stellar mass BH systems, e.g., GRS 1915+105, the half-hour radio periodic oscillations may be connected to the X-ray oscillations with the similar periods,\ucite{62} while in LS I$+61^\circ$303, it was suggested that the radio QPOs could result from multiple shocks in a jet.\ucite{60} In the framework of AGNs, radio QPO models are diverse. The year-long QPOs are generally considered to be the indicator of the orbital motion of binary SMBH systems.\ucite{63} Helical structures in magnetic fields and plasma trajectory are expected in magnetically dominated jets,\ucite{64} so helical motion of blobs or shocks in relativistic jets have been incorporated to interpret periods around hundreds of days in radio, optical or gamma-ray bands in blazars.\ucite{65,66} Recently, the optical and gamma-ray periods around 0.6 day in BL Lacertae were suggested to originate from kink instability in relativistic jets.\ucite{67} The low-frequency QPOs observed in black hole X-ray binaries are thought to possibly be originated from Lense-Thirring precession of inner accretion disk or corona near the fast-spin BH.\ucite{25}

Kink instabilities are a kind of current-driven plasma instability in magnetically-driven jets, which can dissipate significant amount of magnetic energy to accelerate particles. These processes could also produce QPOs due to the quasi-periodic energy release of kink instabilities.\ucite{68} Simulations of the QPO signal based on kink instabilities predict that both flux and linear polarization degree (LP) show modulations with the same period, and their variations are anti-correlated. During the QPO epochs, the flux shows a $\sim 0.2$ s period, while the polarization parameter lightcurves including LP do not show such a QPO period. Then in the QPO phases, there is no correlation between the variations of two parameters (see Figure 3 of \ucite{39}). We therefore also disfavor this mechanism as the origin of the observed subsecond QPOs.

For the helical motion scenario, since steady emission in a helical jet cannot produce any periodic signals,\ucite{65,66} bright clumpy plasma knots in the jet would be required (see the upper panel of Figure 6). This would require the formation of such knots through magnetic reconnection in the twisted magnetic field within the jet, triggered by internal collisions or kink instabilities. If we assume that such blobs move in helical trajectories in the jet, one needs to require that they maintain in a dissipation state without significant change of emission properties across thousands of rounds of helical motion. The subsecond flux modulations also need that the extend of the blobs should be also small enough so that the line of sight would not see the emitter from multiple helices. Otherwise, if there is always strong emission along the line of sight, significant temporal modulation would not be expected. In addition, there also possibly exist multiple bright knots moving in relativistic jets, which could result in the variations and fluctuations of the QPO signals (see wavelet results in Fig. 5). Thus in general jet knot helical motion scenario is complicated, further theoretical work would be requested.

The observed X-ray QPOs of the frequency from $\sim 1-9$ Hz in GRS 1915+105 would be related to the precession of disk or corona. Then it is also naturally expected that the jet base would precess, which would drive the precession of relativistic jet. The radio emission comes from the relativistic jet, thus the 5-Hz radio QPO should be resulted from the special jet dynamics. In the bottom panel of Figure 6, we have presented the possible physical scenario to produce the periodic radio emission based on the Lense-Thirring (LT) precession scenario: the jet direction precesses and emits the electromagnetic radiation. If the BH spin is misaligned with the rotation axis of the disk, the jet base (e.g., the inner accretion disk) around the disk radius $R$ has a LT precession frequency:\ucite{69,70}
\begin{equation}
\nu_{\rm LT}(R)={G\over \pi c^2}{J_{\rm BH}\over R^3},
\end{equation}
where $J_{\rm BH}=aGM_{\rm BH}/c $ is the BH angular momentum and $a$ is the BH spin parameter. Given the observed BH parameters of GRS 1915+105,\ucite{71} i.e. $M_{\rm BH}\sim 12.4 M_\odot$, $a\sim 0.98$, for the typical precession period of $\sim$ 0.2 second, we constrain the typical radius of the jet base as $R\sim 10r_g$, which is consistent with the QPO models based on the LT effect. Thus, in the case of the jet precession, the BH spin direction is misaligned with the rotation axis of the disk (e.g., $\theta_o > 0 $), the jet direction precesses at the frequency of $\sim 5$ Hz with the angle of $\theta_j$, and our viewing angle of this BH system is $60^{\circ}$ to z-axis.\ucite{71}

\vskip 4mm
\fl{1}\includegraphics[width=1.0\columnwidth]{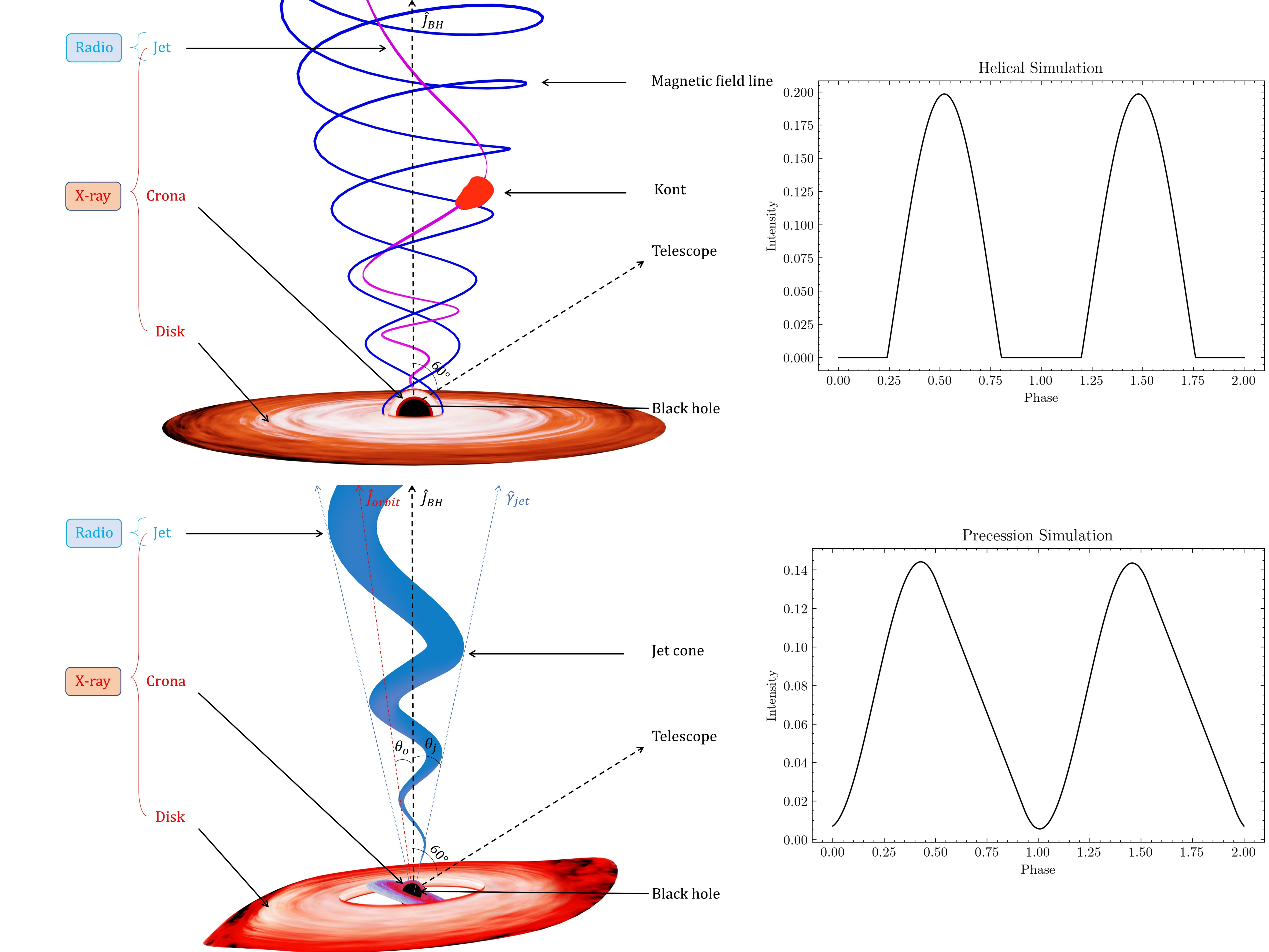}

\vskip 2mm
\figcaption{7.5}{6}{The cartoon models of the black hole with accretion disk and jet in GRS 1915+105 for helical magnetic field case (upper panel) and the precessing jet (bottom panel), respectively. Near the BH, X-rays can be emitted by disk, corona and jet, then only X-ray QPO observations cannot resolve the origin and physics of the QPO. While radio emission comes from jet, the fast variable features in radio observations is the unique diagnosis of the QPO production and physics. {\bf Upper:} The blue lines indicate the helical trajectory of the magnetic field lines along the jet, with the viewing angle of $60^\circ$ for GRS 1915+105.\ucite{71} The knot (or clumpy plasma) in red color moving along the magnetic stream surface (twisted red line) emits the periodic radio wave to the observer. {\bf Bottom:} if the BH spin axis is moderately misaligned with the rotational axis of accretion system (e.g., $\theta_o> 0$), the inner accretion flow as the jet base, precesses and then leads to jet direction precessing. The right embedded panels show the simulated curves from the helix configuration and jet precession, both of which can produce the modulations of observed radio flux due to the relativistic Doppler effects.\ucite{72} }

\medskip

Relativistic jet of GRS 1915+105 has a speed around $0.8-0.99\ c$ based on the radio observations,\ucite{29} the relativistic jet plasma will radiate via synchrotron radiation, and relativistic Doppler effect will affect the observed spectrum. We can derive the observed frequency with the relativistic Doppler effect:\ucite{72}
\begin{equation}
\begin{split}
   \omega = \delta \omega'= \frac{\omega^{\prime}}{\gamma (1-\frac{v}{c} \rm{cos}\theta)},
\end{split}
\end{equation}
where $\omega^{\prime}$ is the frequency of the radiation in the rest frame of the source, $\omega$ is observed frequency, $\delta = [\gamma (1-(v/c) \cos\theta)]^{-1}$ is the Doppler factor, $\gamma$ is the Lorentz factor, $v$ is the speed of the jet, $\theta$ is the angle between line of sight and jet direction.

The observed radio flux in the given frequency range (1.05--1.45 GHz for FAST) will change with the movement of the jet knots or precession due to the Doppler shift. We calculate the relative strength of observed radiations with the QPO phase, where we set the radiation from the jets in the rest frame of the source to be 1. In the right embedded curves of Figure 6, we present the predicted QPO pulse profiles in the cases of helical motion and jet precession scenario. In this picture, we only consider the relativistic jet moving toward to the observer. It is possible that the jet could be produced in the other direction fast-moving away from the observer. The calculations find that radiation of the relativistic jet away from us is relatively low and hard to detect, thus contribution by the jet away from observer to the flux modulation is not considered here because of its low strength.

GRS 1915+105 is the strongly variable X-ray source, and shows flares in the historic records.\ucite{31} Since 2018, GRS 1915+105 unexpectedly started a peculiar low-luminosity state that is an order of magnitude dimmer than  the previous states, with greater hardness ratio in X-rays \ucite{73,74,75} Even though intrinsic dimming is possible, detailed X-ray spectral analyses suggested that the source may have entered an obscured state with the strong absorption (by disk winds or torus in the outer disk part) along the observer's sight \ucite{76,77} due to a large inclination angle of $\sim 60^\circ$.\ucite{71} During the FAST observations on 2021-01-25, the X-ray flux was weak based on both Swift and MAXI observations.\ucite{31,47} The jets revealed by our FAST observations suggests that presently GRS 1915+105 may still have a high accretion rate to power transient relativistic jets. This adds further support to the suggested strong obscuration in X-rays.

{\it Prospectives: } 
With the early FAST observations, we have performed the multiple high-resolution timing monitoring of the microquasar GRS 1915+105. The data analysis procedures including excluding RFIs and calibiration have been well finished for the further science studies. We discovered the fast variations of the radio emissions (including the flux and polarization) from the relativistic jets of GRS 1915+105, specially the 5-Hz radio QPOs are discovered in two instances during the observations in 2021 and 2022. These discoveries at the first time constrain the QPO production site in stellar-mass black hole systems, which should be connected to special dynamics of relativistic jets. The new phenomenon could arise from the precession of a magnetized relativistic jet with a warped accretion disk or possible helical motions of jet blobs.

At present, we only detect the significant radio variability and first radio QPOs in one microquasar GRS 1915+105, and we have no idea of the physics mechanism of producing radio QPOs. Thus we need more FAST observations on the known target GRS 1915+105, and other microquasar candidates. The main science goals are summarized here: (1) Could the radio QPOs be detected in all bright microquasars or during the radio flares of these sources? If radio QPOs exist in some or all microquasars, then there exists some common special dynamics of relativistic jets resulting in the radio QPOs in all BH systems. If the radio QPOs only occur in the special source GRS 1915+105, then we need check why this BH is special and what is the physical origin of the phenomena. (2) Previous multiwavelength studies of BH systems report the flux QPOs in radio, optical and X-ray bands, there are no detailed polarization timing studies of the BHs; then with the high sensitivity observations of FAST and radio polarization ability, we hope to search for the possible polarization quasi-periodic oscillations at the first time in microquasars, which will may directly connect to magnetic field configuration, and instability physics in relativistic jets. These fields are completely new for the BH physics and observations, should be promising to understand the accretion physics, magnetic field and jet production process near BHs.





\textit{Acknowledgements.} We are grateful to the referees for the comments to improve the manuscript. This work is supported by the National Natural Science Foundation of China (Grant No.~12133007), the National Key Research and Development Program of China (Grant No. 2021YFA0718503, and 2023YFA1607901), and the Cultivation Project for FAST Scientific Payoff and Research Achievement of CAMS-CAS. We thank Bing Zhang, Jifeng Liu, Luis Ho, Pengfu Tian, Pei Wang, Xiaohui Sun, Peng Jiang, Ping Zhang, Xiao Chen, Jiashi Chen and Haifan Zhu for the valuable discussions and help to improve the draft.


\end{document}